# FineIBT: Fine-grain Control-flow Enforcement
# with Indirect Branch Tracking

Alexander J. Gaidis*
Brown University
Providence, RI, USA
agaidis@cs.brown.edu

Joao Moreira*
Intel Corporation
Hillsboro, OR, USA
joao.moreira@intel.com

Ke Sun
Intel Corporation
Hillsboro, OR, USA
ke.sun@intel.com

Alyssa Milburn
Intel Corporation
Hillsboro, OR, USA
alyssa.milburn@intel.com

Vaggelis Atlidakis
Brown University
Providence, RI, USA
eatlidak@cs.brown.edu

Vasileios P. Kemerlis
Brown University
Providence, RI, USA
vpk@cs.brown.edu

## ABSTRACT

We present the design, implementation, and evaluation of `FineIBT`: a CFI enforcement mechanism that improves the precision of hardware-assisted CFI solutions, like Intel IBT, by instrumenting program code to reduce the valid/allowed targets of indirect forward-edge transfers. We study the design of `FineIBT` on the x86-64 architecture, and implement and evaluate it on Linux and the LLVM toolchain. We designed `FineIBT`'s instrumentation to be compact, incurring low runtime and memory overheads, and generic, so as to support different CFI policies. Our prototype implementation incurs negligible runtime slowdowns (≈0%−1.94% in SPEC CPU2017 and ≈0%−1.92% in real-world applications) outperforming Clang-CFI. Lastly, we investigate the effectiveness/security and compatibility of `FineIBT` using the ConFIRM CFI benchmarking suite, demonstrating that our instrumentation provides complete coverage in the presence of modern software features, while supporting a wide range of CFI policies with the same, predictable performance.

## 1 INTRODUCTION

Software systems impact virtually every aspect of modern society, ranging from economy and politics to science and education. Today's software, however, is large, complex, and plagued with vulnerabilities that allow perpetrators to exploit it for profit. Of all the different kinds of exploitable software weaknesses, *memory errors* [169]—i.e., bugs that can be (ab)used by attackers, via crafty inputs, to corrupt or leak memory contents [159], or even cause a DoS [179]—have shown to be particularly pernicious to deal with: according to the SANS institute, memory errors dominate the list of the "*top 25 most dangerous software errors*" [150], while vendors like Microsoft and Google attribute ≈70% of the vulnerabilities in their products to memory safety issues [77, 115, 163].

In real-world exploits (against software that is written in memory- and/or type-unsafe languages, like C/C++), attackers primarily trigger *spatial* [122], or *temporal* [123], memory safety violations to tamper with *control data* [93], as these facilitate hijacking the control flow of programs and allow for performing arbitrary code execution [129]. With the adoption of the W^X memory protection policy [130] from contemporary platforms [103, 114], attackers nowadays resort to performing computations via code reuse [32] (e.g., ROP [153], JOP [27, 39], COP [73], JIT-ROP [156], COOP [151]).

*Control-flow Integrity* (CFI) [33] mitigates code-reuse-based attacks (and, in general, exploits that rely on control-flow hijacking) by confining control-flow transfers to benign, allowed (code) locations only. Since the seminal work of Abadi et al. [12], there has been a plethora of CFI schemes proposed, ranging from software-only ones [12, 26, 29, 34, 41, 52, 54, 60, 61, 63, 69, 72, 78, 81, 83, 85, 102, 113, 120, 124–127, 134, 135, 137, 138, 140, 152, 161, 165, 168, 172–174, 184–186, 189], to hardware-assisted [14, 15, 19, 59, 71, 80, 82, 94, 97, 111, 118, 128, 132, 166, 171, 182, 188] and hardware-only solutions [55, 56, 183, 187], which cover different points in the design space regarding effectiveness, coverage, compatibility, *etc.*

CFI is increasingly gaining traction in real-world settings: Microsoft introduced its CFI scheme (i.e., CFG—Control Flow Guard) with Visual Studio 2015 [116], and has since been using it to harden Windows (and more) [160]; Clang/LLVM provides a range of CFI schemes that stem directly from the work of Tice et al. [165], which Google already uses for hardening Android [141] and Chrome [142]; while vendors like Intel and ARM have recently introduced extensions for assisting CFI enforcement (Intel CET, ARM BTI) [48, 98].

Indirect Branch Tracking (IBT) is part of Intel's Control-flow Enforcement Technology (CET) [46, 154], and explicitly protects exploitable *forward-edge* (i.e., via indirect `call`/`jmp` instructions) control-flow transfers from abuse. Recently, support for IBT was added to the Linux kernel, with v6.2 enabling kernel IBT by default on supported (i.e., x86) platforms [107]. Linux has also received preliminary support for IBT in userland [105, 106], as well as in the GNU C library and Binutils [101]; in addition, both Clang/LLVM and GCC support instrumenting code with IBT (via `-fcf-protection`), which popular Linux distributions, like Ubuntu, already enable by default [162]. Although IBT-based CFI hinders code-reuse-based attacks, an attacker who is able to tamper with forward-edge transfers can still "bend" [36] the control flow towards any of the valid-/allowed program (code) points, because the CPU cannot differentiate among different types of IBT-marked code locations. Hence, in terms of effectiveness, IBT-based CFI is equivalent to considering every allowed (indirect) branch target as part of the same (and only) equivalence class. This cheme is analogous to Microsoft CFG, less stringent than Clang-CFI, and bypassable [36, 73].

---

*Joint first authors.

To address this overarching problem, we present `FineIBT`, which aims at providing an *apparatus* for improving the precision of IBT-based CFI, by instrumenting program code to *reduce* the valid-/allowed targets of indirect forward-edge transfers. Advanced static and/or dynamic analyses (hardware-assisted too [59, 71, 80, 82, 166]) that are used for *pruning* the set of allowed targets per indirect control flow transfer, typically via means of enhanced points-to [59, 60, 82, 97, 127, 173], type [78, 102, 120, 165], and class hierarchy [34, 61, 81, 134, 135, 185] inference, are orthogonal to `FineIBT`—we do not require nor preclude any such scheme. `FineIBT` focuses solely on how to enforce a finer-grain CFI policy atop IBT (or ARM BTI).

We study the design of `FineIBT` on the x86-64 architecture, and implement and evaluate it on Linux and the LLVM toolchain. We designed `FineIBT`'s instrumentation to be: (1) *compact*, and thereby incur low runtime and memory overhead(s); and (2) *generic*, so as to support a range of different CFI policies, such as arity-based CFI (i.e., still coarse-grain but more precise than vanilla IBT) [168], strict/relaxed type-based CFI (e.g., à la Clang-CFI; fine-grain) [165], or even finer-grain (and advanced) policies, like context-/path-sensitive CFI [59, 166] and MLTA-based CFI [102], which can reduce the number of valid/allowed targets by up to 98%, or UCT-based (i.e., unique code target) CFI [82] that results in a single target per indirect control flow transfer. Lastly, in antithesis to earlier CFI schemes [12, 161], `FineIBT` is compatible with execute-only memory [45], thus allowing the interplay of `FineIBT` with leakage-resilient code diversification [30, 51, 139]. (Despite leveraging Intel IBT, and the x86-64 architecture, to introduce the concepts behind `FineIBT`, our techniques can boost similar, hardware-based CFI mechanisms, like ARM BTI [98]; see §7.1.)

Our results from evaluating the prototype of `FineIBT` suggest that our prototype incurs a negligible runtime slowdown: ≈0%–1.94% in SPEC CPU2017 and ≈0%–1.92% in real-world applications, like Nginx, Redis, MariaDB, and SQLite, outperforming Clang-CFI (≈0%–7.89% in SPEC CPU2017). In terms of code-size overhead, `FineIBT`'s instrumentation results in moderate (2.27%–19.05%) code-size increase, less than that of Clang-CFI (2.13%–23.21%) and on par with that of similar schemes [33].

In addition, we investigate `FineIBT`'s effectiveness (i.e., security) and compatibility using the ConFIRM CFI benchmarking suite [180]. Our findings demonstrate that `FineIBT`'s nimble instrumentation provides complete coverage in the presence of modern software features (callbacks, exception handling, {load, run}-time code-symbol resolution, *etc.*), while, at the same time, supporting a wide range of different CFI policies (i.e., coarse- vs. fine- vs. finer-grain) with the same, predictable (performance) behavior.

## 2 BACKGROUND AND RELATED WORK

### 2.1 Evolution of Memory-error Exploits

Software that is written in memory- and/or type-unsafe languages, like C, C++, Objective-C, and assembly (ASM), is vulnerable to *memory errors* [169], which enable attackers to *corrupt* or *leak* contents inside the (virtual) address space of victim programs. In real-world exploits, attackers primarily aim for *control data* (e.g., return addresses, function pointers, dynamic dispatch tables; *code pointers*) [93], as these facilitate *hijacking the control flow* of programs and performing *arbitrary code execution* [129].

Historically, arbitrary code execution was achieved via means of *code injection* [84]. However, with the adoption of *non-executable memory* [176] from contemporary platforms [103, 114], and the enforcement of the W^X memory protection policy [130], code injection is nowadays used only in multi-stage exploits [170].

In the present climate, attackers resort to performing arbitrary code execution via means of *code reuse* [32]: i.e., the attacker hijacks the control flow (by tampering with code pointers) and executes benign program code in an "out-of-context" manner. A wide range of code-reuse techniques has been proposed and developed thus far, with the following being the main representatives: ROP [153], JOP [27, 39], COP [73], JIT-ROP [156], BROP [25], SROP [28], COOP [151], CROP [70], AOCR [149], and PIROP [75]. Importantly, BlindSide [76] and SPEAR [109] have recently shown that code reuse is also possible in the *speculative* execution domain.

### 2.2 Control-flow Integrity

The advent of code-reuse-based exploitation has, in turn, prompted the development of a rich set of mitigations [159], with the majority of them designed around the concepts of Control-flow Integrity (CFI) [33], Data-flow Integrity (DFI) [18, 38], automated diversification [23, 40, 91, 95, 145, 155, 177], and memory isolation [90, 93, 143].

CFI was formally introduced in 2005 by Abadi et al. [12], and is effectively a *program shepherding* technique [88, 136]. In particular, under CFI, attackers are allowed to tamper with (i.e., corrupt, inject, swap) code pointers, but control flow transfers are confined to benign, allowed (code) locations only. Since the seminal work of Abadi et al. there has been a plethora of CFI schemes proposed, ranging from software-only ones [12, 26, 29, 34, 41, 52, 54, 60, 61, 63, 69, 72, 78, 81, 83, 85, 102, 113, 120, 124–127, 134, 135, 137, 138, 140, 152, 161, 165, 168, 172–174, 184–186, 189], to hardware-assisted [14, 15, 19, 59, 71, 80, 82, 86, 87, 94, 97, 111, 118, 128, 132, 166, 171, 182, 188] and hardware-only solutions [55, 56, 183, 187]. As expected, this wide body of work covers different points in the design space regarding effectiveness, coverage, compatibility, and target domain.

Nevertheless, the common, underlying thread of each CFI scheme is the existence of a *control flow enforcement mechanism* that effectively "checks" if every computed control flow transfer is on par with a predetermined control flow graph (CFG). Such mechanisms act as *Inlined Reference Monitors* (IRMs) [64, 65] that enforce policies regarding the target(s) of indirect branch instructions.

*2.2.1 Effectiveness.* The effectiveness of a CFI scheme is directly related to its set of allowed targets per indirect control-flow transfer. In general, the more targets allowed, the better it is for an attacker, as there is still enough leeway to perform computations [33]. Earlier CFI solutions provided *coarse-grain* control-flow confinement, ranging from: (a) merely restricting indirect control flow transfers, i.e., via call/jmp/ret in x86 to N-byte (typically N={16, 32}) aligned instructions [113, 152, 182]; to (b) matching every ret instruction with all possible return sites (i.e., code locations following a call instruction) and every (indirect) call/jmp instruction with all possible function-entry points [186]; to (c) improving case (b) above with matching every (indirect) call/jmp instruction with only address-taken function-entry points [189]; to (d) matching every indirect call site with entry points of address-taken functions with the same arity [168]; and (e) everything in-between (a)–(d) [12, 26, 29, 41, 52, 63, 69, 83, 85, 118, 124–126, 132, 138, 140, 174, 184, 188]. Alas, coarse-grain CFI is bypassable [37, 57, 73, 74].

This, in turn, has fueled the development of *fine-grain* CFI solutions [14, 15, 19, 34, 54, 59–61, 71, 78, 80–83, 86, 87, 94, 97, 102, 111, 120, 127, 128, 134, 135, 137, 161, 165, 166, 171–173, 175, 185], which leverage advanced static and/or dynamic analyses (oftentimes hardware-assisted [59, 71, 80, 82, 166]) for further *pruning* the set of allowed targets, typically via means of enhanced points-to [59, 60, 82, 97, 127, 173], type [78, 102, 120, 165], and class hierarchy [34, 61, 81, 134, 135, 185] inference. Fine-grain CFI is not bulletproof either [36, 66, 151]. Attacks against fine(er)-grain CFI take advantage of the inevitable imprecision (e.g., over-approximation) [146] of the aforementioned analyses, thereby "bending" the control flow to code locations that belong to the same *equivalence class* [33, 151].

*2.2.2 Coverage.* Indirect control-flow transfers can be further divided into *forward-* and *backward-edges,* based on the type of transition(s) they correspond to in the CFG. In x86, the former represent branches via (indirect) call/jmp, while the latter via ret. Different CFI schemes provide dissimilar coverage regarding forward and backward edges, ranging from schemes that protect both kinds of transfers [12, 14, 15, 26, 41, 52, 54–56, 59–61, 63, 71, 78, 82, 85, 94, 97, 111, 113, 118, 120, 124–128, 132, 134, 137, 138, 152, 161, 166, 168, 171–175, 182, 183, 186, 188, 189], to ones that focus solely on forward [19, 34, 69, 83, 86, 87, 102, 135, 140, 165, 184, 185] or backward [80, 187] edges. By and large, solutions that provide partial coverage are shown to be easily bypassable [36]. More importantly, as far as backward edges go, Abadi et al. and Carlini et al. have demonstrated the imperative need for effective, high-precision (i.e., fine-grain) backward-edge protection [13, 36], preferably via means of a *shadow stack* [35, 53].

*2.2.3 Compatibility.* Certain CFI schemes operate directly on *binary code*, either via means of static binary analysis [12, 14, 15, 54, 61, 63, 69, 78, 134, 135, 137, 140, 168, 171, 172, 184, 186, 189] and/or specific hardware extensions [41, 55, 56, 59, 71, 80, 94, 118, 128, 132, 183, 187, 188], while others require access to *source code* [19, 26, 34, 52, 60, 81–83, 85–87, 97, 102, 111, 113, 120, 124–127, 138, 152, 161, 165, 166, 173–175, 182, 185]. In general, source-code-agnostic solutions tend to compromise effectiveness, and coverage, in favor of compatibility with commercial off-the-shelf (COTS) software [33]. In antithesis, access to source code typically allows for advanced static (points-to [82, 127, 173], multi-layer type [102, 165], and class hierarchy [81, 185]) analyses that increase the precision (effectiveness) and/or coverage of the respective scheme. In a similar vein, certain CFI solutions are language-agnostic or mostly target software written in C [12, 14, 15, 19, 26, 41, 52, 54–56, 59, 60, 63, 71, 78, 80, 82, 85, 94, 97, 102, 111, 113, 118, 120, 124–128, 132, 137, 152, 161, 165, 166, 168, 171, 172, 174, 175, 182, 183, 186–189], while others are applicable only in C++ [34, 61, 69, 81, 83, 134, 135, 140, 173, 184, 185] or Objective-C [138] codebases.

*2.2.4 Target Domain.* Most CFI schemes are designed to protect binary-only COTS software or userland applications written in C, C++, Objective-C, and/or ASM [12, 19, 26, 34, 41, 55, 56, 59, 61, 69, 71, 78, 80–83, 86, 87, 102, 111, 113, 118, 124–127, 132, 134, 135, 137, 140, 152, 165, 166, 168, 172, 173, 182–189]. However, certain solutions target more niche domains, like OS kernels and hypervisors [52, 60, 63, 85, 94, 97, 120, 161, 174], or software that executes on mobile [54, 138] and embedded (e.g., IoT) [14, 15, 128, 171, 175] devices.

## 2.3 Intel CET

Control-flow Enforcement Technology (CET) [46, 48, 154] is an extension, available in modern Intel CPUs (e.g., 'Tiger Lake'), for assisting with control-flow confinement. CET consists of two parts: (1) a *shadow stack* that concerns backward-edge control-flow transfers; and (2) *IBT* (Indirect Branch Tracking), which targets forward-edge control-flow transfers.

*2.3.1 Shadow Stacks.* For every regular stack, CET adds a shadow stack region, which is indexed via a new register, dubbed %ssp. The shadow stack is a contiguous (expand-down) memory area, whose pages are "marked" accordingly (bit R/W = 0 and D = 1 in the respective page tables), and whose integrity is hardware-enforced: i.e., regular memory stores (executed from *any* ring) are not allowed in shadow stack pages [46, 48].

When enabled, each time a call instruction gets executed, in addition to the return address being pushed onto the regular stack, a *copy* of it is also pushed onto the shadow stack. Conversely, every time a ret instruction gets executed, the return addresses pointed by %rsp and %ssp are popped from the two stacks, and their values are *compared* together. If they differ, an exception (#CP) is raised; else, control is transferred to the intended site.

*2.3.2 Indirect Branch Tracking.* CET introduces a new (4-byte) instruction, i.e., endbr, which becomes the *only* allowed target of indirect call/jmp instructions. In other words, forward-edge transfers via (indirect) call or jmp instructions are *pinned* to code locations that are "marked" with an endbr; else, an exception (#CP) is raised. Moreover, one can exclude certain indirect branches from being confined, by instrumenting them with the notrack prefix. Lastly, endbr instructions are treated as a nop in older CPUs, which lack CET support, or in cases where IBT is turned off [46, 48].

*2.3.3 Speculative Execution.* In addition to the above, CET imposes restrictions on the execution of instructions *speculatively.* Specifically, in the case of ret, the speculative execution of instructions at return sites takes place only if the two return addresses pointed to by %rsp and %ssp match (or predicted by RSB [92, 108]). Similarly, in the case of indirect call/jmp, speculative execution takes place only if the target of the respective branch is an endbr [46, 48].

*2.3.4 System Support.* Currently, the Linux kernel (since v6.2) enables IBT by default [107], and patches exist that provide preliminary userland support for Intel CET in Linux [105, 106], the GNU C library (`glibc`), and Binutils [101]. In addition, both Clang/LLVM and GCC support instrumenting code with IBT (via `-fcf-protection`), which effectively results in adding `endbr` instructions in every *address-taken* function, but, importantly, in every (code) symbol with *non-local* linkage (i.e., every exported function) as well. The latter is important for supporting function calls across dynamic shared objects (DSOs).

Specifically, ELF-based platforms rely on the PLT/GOT (and the dynamic linker/loader; `ld.so`) to support function calls across DSOs [131]. In x86 Linux, every PLT entry includes an indirect `jmp` instruction for branching to the target (code) symbol—potentially located in a different DSO (i.e., `.so` ELF file)—via an accompanying GOT "slot." As GOT entries are writeable addresses (because they have to be updated by `ld.so` *lazily*; i.e., delayed binding), every (code) symbol they may end-up "pointing at" has to be `endbr`-instrumented to prevent attacks that tamper with GOT entries (i.e., GOT overwrite attacks). In other words, every code symbol in a DSO, with non-local linkage, can potentially be the target of a forward-edge transfer via PLT/GOT, and hence needs to be IBT-hardened. (We further discuss this issue in Section 4.4.)

*2.3.5 ABI Changes.* The introduction of IBT mandated an update to the x86-64 PLT format [131] (exemplified in Appendix A.2) to handle address-taken PLT entries—i.e., allow indirect call instructions to target PLT slots—, which typically occur when a symbol with external linkage has its address taken.

Under the standard PLT format, if an IBT-hardened program were to indirectly target a PLT entry, a `#CP` exception would be raised due to the entry having no `endbr` "landing pad." Thus, to accommodate IBT, a new PLT format was introduced that allows PLT entries to be indirectly targeted. The IBT PLT splits each entry across two tables contained in two ELF sections: `.plt.sec` and `.plt`. The former consists of an `endbr` instruction followed by a memory-indirect `jmp` through the GOT, and the latter consists of an `endbr` instruction followed by code to arrange for symbol resolution via the dynamic linker/loader (`ld.so`).

Importantly, the `endbr`s in the `.plt.sec` section are required for indirectly-called external functions, and the `endbr`s in the `.plt` section are required to support delayed binding, as the first time an external function is called, its `.plt.sec` entry will indirectly jump through the GOT to the corresponding `.plt` entry. An unfortunate repercussion of the IBT PLT is the increase of allowed branch targets to include every `.plt.sec` slot (they can be address-taken) and every `.plt` slot (they are targets of `.plt.sec` entries). A thorough description of the IBT PLT is given in Appendix A.3.

## 2.4 Toolchain Support for CFI

*2.4.1 Visual Studio.* Microsoft introduced Control Flow Guard (CFG) with Visual Studio 2015 [116]. CFG (enabled with `/guard:cf`) assigns a bitmap to each process, in which two bits represent 16 bytes of (virtual) memory. In addition, memory locations that correspond to entry points of address-taken functions have their bits asserted in the bitmap.

CFG instruments every forward-edge, indirect control-flow transfer with an IRM (at the call site) that consults the bitmap regarding the target address: if the respective, requisite bits are asserted, then the branch is allowed, as the control will be transferred to a valid entry point; else, an exception is raised.

In terms of compatibility, coverage, effectiveness, and target domain (§2.2), CFG is basically a *compiler-based*, *forward-edge*-only, *coarse-grain* CFI scheme, applicable both to *userland* and *kernel code*. Backward-edge transfers (i.e., via `ret` instructions) are not protected, effectively allowing the attacker to tamper with return addresses. (Microsoft had developed an equivalent of CFG for return addresses, dubbed RFG—i.e., Return Flow Guard—, but never released it due to its many limitations [67].) As far as forward-edges goes, CFG treats every function, whose corresponding bit is asserted in the bitmap, as belonging to the same equivalence class, thereby enforcing a lax CFI policy that is known to be bypassable [73].

*2.4.2 Clang/LLVM.* Clang provides a range of CFI schemes (enabled via `-fsanitize=cfi`) for covering (indirect) control flow transfers via function-pointer dereferences (C/C++), vtable dispatch (C++), and more. The majority of these schemes stem directly from the work of Tice et al. [165], and, specifically, VTV and IFCC. (Interested readers are referred to Appendix A.1 for more information about how Clang-CFI confines forward-edge transfers.)

In terms of compatibility, coverage, effectiveness, and target domain (§2.2), Clang-CFI is a *compiler-based*, *forward-edge*-only, *fine-grain* (type-based) CFI scheme, applicable both to *userland* [22] and *kernel code* [104]. Backward-edge transfers (i.e., via `ret` instructions) are not protected, thereby allowing the attacker to tamper with return addresses. As far as forward-edge confinement goes, Clang-CFI relies on (i.e., requires) *link-time optimization* (LTO) in order to [43]: (1) precisely identify (and analyze the type signatures of) all address-taken functions and function pointers (or `vtable` entries) involved in computed branches; (2) perform symbol mangling (FUNC ⤳ FUNC.cfi, call FUNC ⤳ call FUNC.cfi) in all instances of case (1) above; and (3) instrument indirect call sites.

# 3 THREAT MODEL

## 3.1 Adversarial Capabilities

We assume an attacker who is aiming at hijacking the control flow of (vulnerable) programs by exploiting memory errors [169] in their respective codebases. Specifically, we allow the attacker to (ab)use one or many spatial/temporal [122, 123] memory errors, chain them together (if necessary and/or possible), and, in general, take advantage of them as (and when) needed to construct arbitrary write and arbitrary read exploitation primitives for tampering with code pointers [143]. Formally, we consider an attacker who is able to disclose/corrupt the contents of any readable/writeable memory location in the target (virtual) address space [177], multiple times (if needed), and at arbitrary times during program execution.

Lastly, in the case of memory corruption, we assume that the attacker can overwrite memory content(s) with arbitrary values. In terms of adversarial capabilities, our threat model is on par with the current state-of-the-{art, practice} regarding CFI [33, 43, 116].

## 3.2 Hardening Assumptions

We assume an x86 platform with non-executable memory [103] and a codebase that properly enforces the W^X policy [130]—i.e., code injection [84] is not a viable option for our adversary. We also assume a CPU that supports Intel CET [46, 48, 154] (e.g., 'Tiger Lake' or later). CET transparently protects return addresses (i.e., backward-edge transfers) against corruption, via means of hardware-isolated shadow stacks (§2.3). Note that protecting backward-edge control-flow transfers is imperative (and a precondition) for any fine(er)-grain CFI scheme [36, 167]. Lastly, we assume that IBT (§2.3) is used for confining forward-edge transfers (i.e., via function pointers or dynamic dispatch tables) to code locations marked with endbr. We also assume proper OS/system and toolchain support for Intel CET (§2.3). Turning "off" CET, or disabling IBT/shadow stacks, is not considered within scope.

Given the hardening assumptions above, an attacker can still tamper with forward-edge transfers, and perform code reuse in a {JOP, COP}-like [27, 39, 73] fashion: with IBT, the resulting (coarse-grain) CFI scheme can, at most, confine forward-edge transfers to (all) address-taken function-entry points (§2.2)—and, to make matters worse, the above set includes every (code) symbol with non-local linkage (e.g., every exported function) as well [119] (§2.3), resulting in a loose CFI policy that provides limited protection [73]. All other, standard hardening features (e.g., ASLR [68], stack-smashing protection [50]) are orthogonal to the scheme(s) we propose; we do not require nor preclude them. The same is also true for less-widespread mitigations, like code randomization/diversification [91, 95, 177] and protection against data-only attacks [143].

## 4 DESIGN

### 4.1 Overview

*4.1.1 Synopsis.* FineIBT is a new CFI enforcement scheme that focuses on performance, support for fine-grain policies across different architectures, and compatibility with other defenses, such as execute-only memory. It achieves this by augmenting existing hardware-based, forward-edge CFI schemes with compact IRMs to enforce fine-grain policies in lieu of what would typically be coarse-grain policies. Careful consideration was given to the design of FineIBT's IRM to ensure optimal performance (§4.3) and flexibility. Further, FineIBT securely supports cross-DSO function calls (§4.4) and the ability to debloat libraries at runtime, by refining equivalence classes of indirect branch targets (§4.5).

*4.1.2 Approach.* FineIBT is a CFI enforcement scheme that is designed for increasing the effectiveness of hardware-based, forward-edge control-flow confinement mechanisms, including, but not limited to, IBT (§2.3). Under IBT (§3), an attacker who is able to tamper with forward-edge transfers can still "bend" [36] the control flow towards any of the valid/allowed function-entry points marked with endbr, because the CPU cannot differentiate among different types of endbr-marked code locations. IBT-based CFI is analogous to Microsoft CFG, less stringent than Clang-CFI (§2.4), and prone to COP-style code reuse [36, 73].

FineIBT aims at providing the apparatus for improving the precision of the enforced CFI policy, when IBT, is used, by instrumenting the respective code with IRMs that reduce the valid/allowed targets of indirect forward-edge transfers. Advanced static and/or dynamic analyses (hardware-assisted too [59, 71, 80, 82, 166]) that are used for pruning the set of allowed targets per indirect control flow transfer, typically via means of enhanced points-to [59, 60, 82, 97, 127, 173], type [78, 102, 120, 165], and class hierarchy [34, 61, 81, 134, 135, 185] inference, are orthogonal to FineIBT. FineIBT focuses solely on how to enforce (in an effective and performant manner) a finer-grain CFI policy, which can be generated by any (program analysis) technique, atop IBT. In what follows, we assume that IBT/FineIBT is added to target code via means of compiler-based instrumentation (§2.3). However, this is not a hard requirement, as recent advances in binary code rewriting (i.e., frameworks like Egalito [178], RetroWrite [58], BinRec [20]) allow retrofitting binary-only software with support for IBT/FineIBT.

We designed our IRM code to be: (1) *compact*, and thereby incur negligible runtime and memory overhead; and (2) *generic*, so as to support a range of different CFI policies, such as arity-based CFI (i.e., still coarse-grain but more precise than vanilla IBT) [168], strict/relaxed type-based CFI (e.g., à la Clang-CFI; fine-grain) [165], or even finer-grain (and advanced) policies, like context-/path-sensitive CFI [59, 166] and MLTA-based CFI [102], which can reduce the number of valid/allowed targets by up to 98%, or UCT-based (i.e., unique code target) CFI [82] that results in a single target per indirect control-flow transfer. Crucially, the IRM code of FineIBT is *agnostic* to the exact policy that is enforced. In other words, despite whether we are enforcing, say, arity- or MLTA-based CFI, the respective IRM code is *exactly* the same, hence completely decoupling the associated runtime and memory overhead(s) from the effectiveness (i.e., strictness) of the selected policy. Lastly, in antithesis to earlier CFI schemes [12, 161], the IRM code of FineIBT is compatible with execute-only memory [45], allowing the use of FineIBT with leakage-resilient code diversification [30, 51, 139].

### 4.2 Foundational FineIBT Instrumentation

Listing 1 illustrates FineIBT's IRM code in its most basic form. Function main contains an indirect forward-edge transfer (ln. 4), while func0 and func1 are address-taken functions—and hence potential targets of the indirect call instruction in main. Based on the above, IBT instruments the *prologue* of func0 and func1, with endbrs, confining the indirect branch in main (ln. 4) to func0 (ln. 8) or func1 (ln. 15). In contrast to state-of-practice CFI (e.g., Clang-CFI [165], Microsoft CFG [116]), FineIBT instruments both the callers and the callees involved in indirect forward-edge transfers. Specifically, indirect call sites (i.e., callers) are instrumented with a single mov instruction that loads an integer value to a general-purpose, volatile (e.g., caller-saved) register. In the example above, the indirect call instruction in main (ln. 4) is instrumented with a mov instruction that loads the value 0xc00010ff to register %eax (ln. 3). Moreover, every function, say, FUNC, which is instrumented with an endbr has its prologue code "moved" under the symbol FUNC_entry that has the same visibility as FUNC (ln. 13, ln. 20).

```
 1  main:                     /* caller */
 2  ...
 3    mov    $0xc00010ff, %eax   /* SID = 0xc00010ff */
 4    call   *%rcx
 5  ...
 6    call   func1_entry
 7  ...
 8  func0:                    /* callee */
 9    endbr64
10    sub    $0xc00010ff, %eax   /* SID = 0xc00010ff */
11    je     func0_entry
12    hlt
13  func0_entry:
14  ...
15  func1:                    /* callee */
16    endbr64
17    sub    $0xbaddcafe, %eax   /* SID = 0xbaddcafe */
18    je     func1_entry
19    hlt
20  func1_entry:
21  ...
```

**Listing 1: Basic FineIBT IRM code (caller-callee) in x86-64.**

Next, FineIBT inserts three instructions between the endbr and the respective function prologue: a sub instruction, which subtracts a value from the general-purpose volatile register (ln. 10, ln. 17); a je (direct conditional branch) instruction (ln. 11, ln. 18); and a (single-byte) hlt (0xf4) instruction (ln. 12, ln. 19). Lastly, direct call targets are replaced with their *_entry counterparts, effectively bypassing completely both the instrumentation of IBT and FineIBT (ln. 6; call func1 ↝ call func1_entry)—this way FineIBT has *zero* impact on direct call instructions.

We also refer to the integer value loaded-in/subtracted-from %eax (ln. 3, ln. 10/ln. 17) as *set ID* (SID). In particular, SIDs are used for: (a) *assigning* (address-taken) functions to equivalence classes; and (b) *associating* indirect call sites with their equivalence class. In the example above, func0 is mapped to an equivalence class whose SID = 0xc00010ff (ln. 10), func1 is mapped to a class whose SID = 0xbaddcafe (ln. 17), and the indirect call (ln. 4) is associated with the same equivalence class of func0.

Based on the above, FineIBT bolsters IBT as follows: first, the SID 0xc00010ff is loaded in %eax (ln. 3), right before executing the indirect call (ln. 4); because of IBT, the indirect call (in main) is only allowed to transfer control to func0 (ln. 8) or func1 (ln. 15). Next, if the indirect call rightfully branches to func0, then the sub instruction (ln. 10) will subtract the SID value 0xc00010ff from %eax, which, in turn, will result in the subsequent branch (ln. 11, je func0_entry) to be taken, skipping the hlt instruction and reaching the actual prologue of func0; else, if the indirect call branches to func1, IBT will still allow the corresponding control flow transfer to take place (i.e., func1 contains an endbr), but the subsequent sub-je instruction pair will result in executing the hlt instruction, as the SID of func1 (0xbaddcafe) does not match the one loaded in %eax (0xc00010ff, ln. 3).

```
 1  .func0_fineibt_coldpath:
 2  ...                       /* arg0, ..., argn */
 3    call   __fineibt_chk_fail@PLT
 4  func0:                    /* callee */
 5    endbr64
 6    sub    $0xc00010ff, %eax   /* SID = 0xc00010ff */
 7    jne    .func0_fineibt_coldpath
 8  func0_entry:
 9  ...
```

**Listing 2: Custom error-handling FineIBT IRM code.**

### 4.2.1 Effectiveness.
The net effect of the above is the *refinement* of the *coarse-grain* policy enforced by (the underlined hardware-based CFI scheme) IBT, which allows both func0 and func1 to be valid targets (of the indirect call instruction in main; ln. 4), to a more *fine-grain* (software-assisted) one that includes func0 only. Note that this refinement process is controlled *solely* by the selection of the respective SID values. The user of FineIBT can select the SID values accordingly, in order to implement stringent CFI policies atop IBT, like arity- [168], type- [165], MLTA-based [102] CFI, *etc.* Lastly, note that SIDs are 4 bytes (32-bit) long, effectively allowing for more than 4 billion equivalence classes.

### 4.2.2 Compatibility.
In contrast to various state-of-practice CFI schemes (e.g., Clang-CFI), which *mandate* LTO support, FineIBT imposes *zero* additional requirements to the process of building (and hardening) a particular codebase with IBT—unless LTO is required by the underlying program analysis technique(s) used for selecting SIDs. This allows every translation unit to be instrumented independently and "as is," easing the adoption of FineIBT. Moreover, by having the IRM logic "split" between caller-callee code, FineIBT allows for incremental deployment, as certain applications (i.e., callers) can be hardened first but still interoperate with unhardened libraries (i.e., callees)—at the expense of reduced protection, of course. Note that the use of LTO has various adverse effects [17]: it mixes together program and library code, into one big "blob", preventing library code from being updated independently (i.e., without recompiling everything).

### 4.2.3 Target Domain.
The IRM code of FineIBT can be used verbatim in every setting that IBT is available at: e.g., userland code, kernel code [121], or even enclave (SGX) code (§2.3). In this work, we study FineIBT in the context of userland code.

## 4.3 Performance Considerations

The performance (overhead) of FineIBT is directly related to the size and structure of the IRM code.

### 4.3.1 CPU Front-end and I-Cache.
We designed the SID-based checking logic of FineIBT (§4.2) so that it is friendly to the CPU front-end (i.e., instruction fetch and decoding unit) and instruction cache (I-Cache) [47]. Specifically, in x86-64, the IRM code of FineIBT consists of 5 bytes per every indirect call site (mov $SID, %reg; caller), and 8 bytes per every endbr-marked code location that is instrumented accordingly (sub-je-hlt; callee)—12 bytes, in total, if we also consider the size of the endbr instruction (4 bytes).

The additional pressure of such a caller-callee (i.e., split) instrumentation scheme in the CPU front-end, and I-Cache, is (considerably) less than a scheme like Clang-CFI, which requires 21 bytes of additional code per indirect call site, as well as 8 bytes per trampoline entry (see Listing 5 in Appendix A.1). (Microsoft CFG is also caller-heavy [24].)

*4.3.2 Instruction Selection.* Compilers typically generate code that uses the xor instruction/idiom to implement comparisons that test for (non-)equality [62]. This would nominally result in an xor instruction in ln. 10/17 of Listing 1, which would perform the same SID-based check with the instrumentation described in Section 4.2. However, during the preliminary evaluation of FineIBT, we noticed that xor-je is an instruction pair that is not *macro-fused* (i.e., merged to a single *μop* [44]), in antithesis to sub-je, which does get macro-fused in modern Intel CPUs. sub-je leads to (1) increased instruction decoding throughput and (2) lower execution latency [47], and we hence force FineIBT to generate this instruction pair (or sub-jne) in the IRM code. Additionally, we opted to use sub over cmp as sub also clears the SID register, preventing SID reuse in the event that an indirect branch is not prefixed with a SID-set instruction.

*4.3.3 Custom Error-handling Code.* Whenever a runtime policy violation occurs, FineIBT results in executing the (single-byte) hlt instruction (§4.2), which, in turn, causes a #CP exception to be raised. If terminating the victim process abruptly is not preferable, FineIBT also supports executing a custom violation/error handler. However, invoking such a handler requires (at least) 5 bytes, for the encoding of a (direct) call instruction, plus additional bytes for argument passing (should the error handler function receive any arguments). Replacing a 1-byte (hlt) instruction, in the IRM code of FineIBT, with a 5-byte instruction stream (at minimum), incurs additional, unnecessary pressure on the CPU front-end and I-Cache, as it pollutes hot paths with instructions that will never execute under normal circumstances.

To alleviate the performance impact of the above, we support an alternative IRM instrumentation, shown in Listing 2, which covers cases of error-handling code that require multiple bytes. Specifically, the multi-byte instruction stream that replaces hlt (ln. 2–ln. 3) is placed outside the body of the respective function, under the label *_fineibt_coldpath. In addition, the sub-je instruction pair is replaced with sub-jne, which performs the SID-based check, as before, and branches to *_fineibt_coldpath in case of a violation; else, the control falls-through to *_entry. Note that sub-jne is also macro-fused, while the symbol of the corresponding function (e.g., func0) retains Intel's alignment recommendation of being at or near a 16-byte boundary [47]. Lastly, we choose to place *_fineibt_coldpath "above" the function-entry point so as to use the compact, 2-byte encoding of jne.

## 4.4 Security Considerations

The x86-64 ABI [21] specifies a new PLT format to support IBT, which grossly increases the set of allowed branch targets. (Interested readers are referred to Appendix A.3 for more details regarding the IBT PLT/GOT mechanism.)

```
1   PLT0:     shl      $0x20, %rax
2             or       $SID, %rax
3             pushq    GOT+8(%rip)        /* GOT[1] */
4             jmp      *GOT+16(%rip)      /* GOT[2] */
5             nopw     %cs:0x0(%rax,%rax,1) /* PAD  */
6   ...
7   PLT4:     endbr64
8             cmp      $SID, %eax
9             pushq    $0x3
10            xchg     %ax, %ax           /*  PAD  */
11            je       PLT0
12            hlt
13            nopw     0x0(%rax,%rax,1)   /*  PAD  */
14  ...
15  FPLT4:    mov      $SID, %eax
16            jmp      *fsym4@GOT(%rip)   /* GOT[6] */
17            nopl     0x0(%rax,%rax,1)   /* GOT[6] */
18  ...
19  ATFPLT4:  endbr64
20            sub      $SID, %eax
21            je       FPLT4
22            hlt
23            data16 nopw %cs:0x0(%rax,%rax,1) /*PAD*/
24            nopl     (%rax)             /*  PAD  */
```

**Listing 3: FineIBT PLT (x86-64 ABI-compatible).**

*4.4.1 FineIBT PLT.* We propose an improved PLT scheme to be used with FineIBT that (a) reduces the IBT PLT over-approximation by up to a factor of 2 (see §A.3), (b) transparently supports both lazy and eager binding, (c) allows for SID-based checking, and (d) is compatible with the x86-64 ABI [21]. (Note that the latter is important as it enables the use of existing tools, like the GNU Binutils, with FineIBT-hardened ELF files.)

Listing 3 illustrates FineIBT's new, improved PLT format, which consists of three tables: (1) PLT0–PLTn, mapped at the .plt ELF section; (2) FPLT1–FPLTn, mapped at the .plt.fineibt ELF section (new); and (3) ATFPLT1–ATFPLTn, mapped at the .plt.atfineibt ELF section (new). Cross-DSO function calls (via the PLT) are linked with entries in .plt.fineibt. So, if the current DSO needs to, say, invoke fsym4, which is located in a different .so, then the respective (direct) call will be as follows: call fsym4@FPLT—i.e., a direct branch to symbol FPLT4 (ln .15).

.plt.fineibt is analogous to .plt.sec of IBT PLT (see §A.3; i.e., it includes an entry for each external symbol of type FUNC), but with the following differences: first, entries in .plt.fineibt do not contain endbr instructions, as they are explicitly targeted by direct call instructions only; second, given that every slot in .plt.fineibt includes a memory-indirect jmp, via GOT, each such entry includes the caller-part of FineIBT's IRM code (ln. 15, ...).

As before, the memory-indirect jmp instructions in .plt.fineibt will either transfer control to the corresponding symbols in foreign DSOs, or to an entry in .plt (lazy binding)—every entry in .plt.fineibt has an associated entry in .plt (FPLTn ↝ PLTn; ln. 7, ...). However, as .plt entries can be the target of indirect jmp instructions (from `.plt.fineibt`), they include the callee-part of FineIBT's IRM code. Lastly, every address-taken PLT slot has an associated entry at `.plt.atfineibt` for handling indirect control-flow transfers to it. Such entries are targets of indirect `call` instructions, and include the callee-part of FineIBT's IRM code; after performing FineIBT's SID-based check, they branch to the corresponding entry in `.plt.fineibt`, which, in turn, transfers control to the target symbol in a foreign DSO, or the respective `.plt` slot.

When an unresolved symbol is encountered for the first time, in the case of lazy binding, the dynamic linker/loader is invoked from PLT0 to resolve and indirectly branch to the target symbol. This requires two SIDs be passed to the dynamic linker: one to verify the callee of the resolver code (i.e., PLT0) and another to verify the callee of the resolved symbol (i.e., the resolver code itself). We solve this problem by preserving the symbol's SID in its respective `.plt` slot (using `cmp` instead of `sub` for the SID check; ln. 8) before jumping to PLT0, where the preserved and new SID values are made to share the upper and lower 32 bits of `%rax`, respectively (ln. 1–2). The dynamic linker can then use the lower 32 bits of the register for its SID check and pass the upper 32 bits of `%rax` to the resolved symbol for its own, subsequent SID check.

Note that although there is a 1-1 correspondence between slots in `.plt.fineibt` and `.plt`, only a subset of slots in `.plt.fineibt` (if any) have associated entries in `.plt.atfineibt`—i.e., the ones that are truly address-taken in the current DSO, effectively *debloating* the IBT PLT from excessive `endbr` instructions that grossly increase the set of allowed branch targets (see §A.3). Also, because of the above table relationships, the SID value in every FPLT slot should be *identical* to the SID value in its associated PLT slot (and ATFPLT slot, if one exists), irrespective of the *sensitivity* of the SID-selection mechanism/analysis. Finally, note that the proposed PLT format supports both lazy and eager binding seamlessly.

*4.4.2 Compact FineIBT PLT.* If RELRO (relocation read-only; `-z relro` or `-Wl,-z,relro`) [147] is configured to operate in "full" mode (i.e., by enabling `BIND_NOW` as well; `-z now` or `-Wl,-z,now`), the dynamic linker/loader will resolve all GOT entries at load time and write-protect `.got`, preventing an attacker from tampering with the GOT mechanism completely (and the binding process, in general). FineIBT can take advantage of the above, and emit an alternative, *compact* PLT as follows.

First, the `.plt` section is omitted entirely, as full RELRO effectively results in eager binding (PLT0–PLTn will never be exercised). Second, entries in `.plt.fineibt` are not instrumented with the caller-part of FineIBT's IRM code, while the respective memory-indirect `jmp` instructions are prefixed with `notrack` [46], allowing them to branch to (code) locations that are not necessarily `endbr`-instrumented. Given that GOT is immutable, `ld.so` can safely link each PLT-related GOT slot regarding foreign symbols with their `*_entry` counterparts, effectively bypassing the callee-part of FineIBT's IRM code. This approach has the added benefit of being slightly more performant than the original one (§4.2), as it avoids the unnecessary execution of SID-based checks, when full RELRO is enabled. Lastly, both our PLT schemes can co-exist in the same address space and interoperate seamlessly. The compact PLT scheme, however, requires an `ld.so` that knows how to handle FineIBT-related `*_entry` symbols.

## 4.5 IBT-instrumentation Elision

As discussed in Section 4.4.1, every DSO (code) symbol with non-local linkage will be `endbr`-instrumented (by IBT), and, consequently, IRM-protected by FineIBT as well, as it may be linked-with a PLT/GOT entry during load time. However, given an address space instantiation, every such symbol will not end up being linked-with PLT/GOT slots. For instance, glibc exports more than 1.5K symbols of type FUNC (v2.31), but only a *subset* of these will end up in the PLT/GOT slots of other DSOs or the main application executable. (In fact, this is exactly what debloating systems rely on to remove redundant code from shared libraries [16, 17, 144].)

*4.5.1 NOPout.* FineIBT includes a load-time feature, dubbed NOP-out, which aims at *eliding* unnecessary `endbr` instructions from DSOs, effectively reducing even further the set of allowed (indirect) branch targets, irrespective of the underlying sensitivity of the SID-selection mechanism/analysis. NOPout is meant to work best with eager binding, but, in principle, the scheme can support lazy binding as well (albeit with additional runtime overhead).

Specifically, FineIBT notes every (a) non-address-taken, (b) non-local, and (c) FineIBT-protected symbol of type FUNC, in a special ELF section (of type SHT_NOTE), dubbed `.plt.nopout`. During load time, once the dynamic linker/loader has resolved *every* PLT/GOT entry (for *all* DSOs loaded in the address space), `ld.so` may proceed to implement NOPout as follows: for each entry in `.plt.nopout` (for all loaded DSOs), `ld.so` checks if the respective entry is linked-with some PLT/GOT slot; if not, then the `endbr` instruction, at the prologue of the corresponding function, can be elided, via means of code patching. Note that at this point, the execution of the main application has not yet commenced, allowing us to avoid the culprits of live code patching [177]. `endbr` instructions are replaced with a 4-byte nop, while in the case of compact FineIBT PLTs (§4.4.2), NOPout will effectively result in an address space instantiation where only address-taken functions are IBT- and FineIBT-protected, which is also the *minimum* set of code locations that need to be hardened by CFI for the application to function properly.

*4.5.2 Dynamically-loaded DSO Support.* If NOPout is in effect, and a DSO is *dynamically-loaded*, e.g., via dlopen, then certain `endbr` instructions (in other DSOs) that were previously elided, may have to be placed back—the newly-loaded DSO may link them in its own PLT/GOT. Similarly, if `dlsym` is used to *dynamically address-take* a function, then, again, if the `endbr` instruction in the prologue of that function was previously elided, it now needs to be placed back.

However, placing back `endbr` instructions, while the application is running, in a *safe* manner, requires special care; NOPout handles such cases as follows. First, it identifies the memory page (4KB) inside of which an `endbr` (or more) instruction(s) should be placed back. Next, it creates a temporary copy of that page using a different location in the address space, which is mapped as non-executable (RW-). Then, the respective 4-byte nop instruction(s) are replaced with endbr(s) in that copy, which is then mapped as read-only (R--). After that, NOPout compares the contents of the page-copy with the original page to verify that the only difference(s) stem from nop ⤳ endbr. Lastly, if the above is true, it remaps the newly-patched code over the original one, discarding at the same time the latter and the temporary mapping.

Note that during this task, `NOPout` processes `.plt.nopout` entries directly from the respective ELF section, which is mapped as read-only to prevent attackers from tampering with the live patching capabilities of `FineIBT`.

## 5 IMPLEMENTATION

### 5.1 System Support

We used Intel's official patches for the Linux kernel (v5.13) [49], which provide support for Intel CET in userland. We slightly edited the patchset to enable diagnostic messages during benchmarking.

### 5.2 FineIBT Toolchain

We implemented `FineIBT` as a set of modifications to the LLVM toolchain (v12). Our prototype consists of (a) ≈700 C++ LOC added to the LLVM compiler (`llvm-{gcc, g++}`) and (b) ≈200 C++ LOC added to the LLVM linker (`lld`) to facilitate the instrumentation, and optimizations, outlined in Section 4. (Note that our design is not LLVM-specific; `FineIBT` can be easily ported to GCC or other compiler toolchains/frameworks.) As far as (a) goes, we extended LLVM's `MachineFunctionPass` [100] to instrument function prologues and call sites, accordingly (i.e., add `FineIBT` IRM code; see §4.2 and §4.3); regarding (b), we added support for regular and compact `FineIBT` PLTs in `lld` (see §4.4). Lastly, we used `%r11` in `lld` to implement SID-based checking—the x86-64 ABI [21] defines `%r11` as a scratch register, simplifying the handling of ASM code.

#### 5.2.1 CFI Policies.
We implemented the following CFI policies (to demonstrate the ability of our flexible enforcement mechanism to support different policies, ranging from coarser- to finer-grain): (1) arity-based CFI [168]; (2) strict/relaxed type-based [165]; and (3) MLTA-based [102]. In the case of (1) and (2), we rely on LLVM type metadata to create the respective equivalence classes and assign SIDs. In the case of (3), we added support to `FineIBT` for leveraging the output of TypeDive [164] (i.e., an LLVM-based MLTA framework by Lu and Hu [102]) and assign functions to equivalence classes based on the caller-callee pairs reported by the tool.

#### 5.2.2 NOPout.
We implemented `NOPout` in ≈800 C LOC. Our prototype consists of a DSO (`libnopout.so`) that hooks the dynamic linker/loader (`ld.so`), via preloading, for patching program code accordingly (§4.5). We decided not to integrate `NOPout` in `ld.so` to allow for seamless experimentation with different dynamic linker/-loader implementations.

## 6 EVALUATION

We performed all our experiments on a host equipped with a 4-core Intel Core i7-1185G7 3GHz CPU and 16GB of (LPDDR4) RAM, running Void Linux [11]. All benchmarked applications, including their (DSO) dependencies, were instrumented/hardened using the prototype discussed in Section 5, built as position-independent (`-f{PIC, PIE}`, `-pie`), and linked-with `musl libc` [7]—vanilla `glibc` cannot be built (yet) with Clang/LLVM because of the extensive use of {GCC, GNU}-specific features [112]. Moreover, all programs were linked with full RELRO (`-z relro -z now` or `-Wl,-z,relro, -z,now`) and use the compact `FineIBT` PLT (§4.4.2); a decision largely motivated by `musl`'s lack of support for lazy binding [5].

**Table 1: SPEC CPU2017 (SPECspeed Integer) results.**

| Benchmark | IBT | FineIBT (§4.2) | FineIBT (§4.3) | Clang-CFI |
|---|---|---|---|---|
| `600.perlbench` | 1.46% | 1.51% | ≈0% | 1.17% |
| `602.gcc` | 0.03% | 0.12% | 0.05% | 0.6% |
| `605.mcf` | 0.51% | 3.67% | 1.94% | 2.04% |
| `620.omnetpp` | ≈0% | 0.91% | ≈0% | 6.57% |
| `623.xalancbmk` | ≈0% | 0.99% | ≈0% | 7.89% |
| `625.x264` | 0.05% | 0.29% | ≈0% | 0.93% |
| `631.deepsjeng` | 0.03% | 0.17% | ≈0% | ≈0% |
| `641.leela` | ≈0% | ≈0% | ≈0% | ≈0% |
| `657.xz` | ≈0% | ≈0% | ≈0% | 0.06% |

Finally, dynamic frequency and voltage scaling (DVFS, Turbo Boost) was disabled, while the CPU was configured to constantly operate at 3GHz in the C0 C-state, in order to minimize benchmarking noise and facilitate reproducibility.

### 6.1 Performance

#### 6.1.1 SPEC CPU2017.
We used SPEC CPU2017 [31] to assess the runtime slowdown that is incurred by `FineIBT`, as well as the impact of the optimizations outlined in Section 4.3. We also compare and contrast `FineIBT` to IBT and Clang-CFI. More specifically, we used the SPECspeed 2017 Integer suite that contains 9 C/C++ applications and a Fortran program, which we excluded from our benchmarks. Table 1 summarizes our findings. The 9 C/C++ applications (col. 1) were built natively, instrumented with IBT (col. 2), hardened with the *basic* IRM code illustrated in Listing 1 (col. 3) and the *optimized* IRM code shown in Listing 2 (col. 4), as well as protected with Clang-CFI (col. 5). Results (avg. over 10 runs, using the `ref` workload) are reported as percentages atop the uninstrumented baseline. (≈0% corresponds to < 0.01%.)

Notably, the `-fsanitize-cfi-cross-dso` flag was used when building libraries and executables with Clang-CFI to allow the CFI scheme to apply across DSO boundaries; otherwise, cross-DSO calls are implemented as if the callee did not have Clang-CFI support [43]. Additionally, `600.perlbench`, `602.gcc`, `605.mcf`, `620.omnetpp`, and `657.xz` failed to run with Clang-CFI due to mismatched types—e.g., in `602.gcc_s`, a function pointer `rtx (*insn-_gen_fn) (rtx, ...)` is used to target functions without variable arguments, such as `rtx emit_move_insn_1 (rtx, rtx)`, which results in a segmentation fault when invoked since there is no corresponding, type-matched entry in the enforcement mechanism's jump tables. We patched these 5 benchmarks to get type agreement via casting or changing function signatures. Our changes do not affect the behavior or functionality of the benchmarks, and the patches were applied across all builds of SPEC (i.e., the uninstrumented baseline, IBT, the two `FineIBT` variants, and Clang-CFI).

IBT incurs negligible runtime overheads (0%–1.46%), which is on par with Intel's previously-reported results [12]. The basic `FineIBT` IRM code incurs a moderate slowdown (0%–3.67%), whereas the optimized one follows closely the performance of IBT, incurring only 0%–1.94% overhead (+1.43% atop IBT in the worst case).

**Table 2: Real-world application results.**

| Application | IBT | FineIBT (§4.3) |
|---|---|---|
| Nginx (1KB) | ≈0% | ≈0% |
| Nginx (100KB) | 0.77% | 1.92% |
| Nginx (1MB) | ≈0% | 0.11% |
| Redis (GET) | 0.39% | 1.17% |
| Redis (SET) | 0.39% | 1.17% |
| MariaDB | 0.55% | 0.60% |
| SQLite | ≈0% | 0.36% |

Both `FineIBT` schemes are friendly to the CPU front-end and I-Cache, and exploit macro-fusion, resulting in minimal slowdowns (§4). However, the latter (col. 3, optimized IRM scheme) removes the error-handling code completely from hot paths, resulting in even lower overheads (atop IBT).

In general, as is the case with IBT, the runtime slowdown of `FineIBT` is (positively) correlated with the number of address-taken functions and indirect `call` sites. Furthermore, the (code) size of indirectly-invoked functions affects the runtime overhead too— larger functions tend to amortize the cost of `FineIBT`'s IRM code. In antithesis, Clang-CFI's overhead ranges between 0%–7.89%, which demonstrates the benefits of our split-instrumentation (i.e., caller-callee), hardware-assisted CFI scheme. Lastly, note that the above overheads are orthogonal to the underlying CFI policy—i.e., no matter how strict or lax the enforced CFI scheme is, `FineIBT` will always incur the same overhead(s), allowing for predictable runtime (performance) behavior.

#### 6.1.2 Real-world Applications.
In addition to SPEC CPU2017, we also used the following set of real-world applications to further investigate the runtime behavior of `FineIBT`: Nginx (v1.20.2) [8], Redis (v6.0.9) [9], MariaDB (v10.5.10) [6], and SQLite (v3.38) [10]. Results (avg. over 20 runs) are reported as percentages atop the uninstrumented baseline. (≈0% corresponds to < 0.01%.) Table 2 summarizes our findings—`FineIBT` (col. 3) corresponds to the optimized IRM variant discussed in Section 4.3.

**Nginx.** We used the `wrk` [4] benchmarking tool to generate HTTP requests, continuously for 1 minute, for three different file sizes: 1KB, 100KB, and 1MB. (The served files were filled with random data.) Both `nginx` and `wrk` ran on the same host, connected via the loopback (`lo`) virtual network interface to minimize I/O latency and increase CPU utilization. Furthermore, `wrk` was configured to use 4 execution threads, each making 128 (simultaneous) HTTP reqs, while `nginx` used 4 worker threads in the case of 1KB and 100KB file reqs and 2 worker threads in the case of 1MB file reqs. We used the above settings to saturate our CPU (max. utilization) and avoid I/O masking the overhead(s) of `FineIBT`. IBT incurs negligible throughput (tput) degradation (0%–0.77%), while `FineIBT`'s tput degradation ranges between 0%–1.92% (+1.15% atop IBT in the worst case). Note that these are worst-case overheads, as if the CPU is not saturated the impact of `FineIBT` is barely measurable.

**Redis.** We used the `memtier_benchmark` [148] tool to generate a stream of SET and GET reqs for a 32-byte object, on a 1 : 10 ratio, continuously for a 1-minute duration. Moreover, `memtier_benchmark` was configured to use two execution threads, each making (up to) 128 simultaneous reqs, while `redis-server` used a single worker thread. Both processes executed on the same host and performed I/O over `lo`. (Again, these parameters resulted in max. CPU utilization.) IBT incurs negligible tput degradation (0.39%), while `FineIBT`'s tput degradation is 1.17% (+0.78% atop IBT).

**MariaDB.** We used sysbench [3] to generate an online transaction processing (OLTP) workload (i.e., the `oltp_read_write` benchmark), and tuned `mariadb` according to the project's recommendations [110]. We ran `oltp_read_write`, continuously for 5 minutes, performing transactions on a single table of 2-million rows (≈500MB of data). sysbench was configured to use 64 execution threads, while `mariadb` used 6 worker threads to saturate the CPU. Both processes executed on the same host and performed I/O over `lo`. FineIBT (IBT) incurs negligible tput degradation: 0.60% (0.55%).

**SQLite.** We used the SQLite Speedtest benchmark [157], which stress-tests `sqlite` and reports the run time for performing a series of DB operations. The benchmark was configured to use an in-memory database to prevent I/O from masking overhead and default settings for all other options. Furthermore, Speedtest is distributed as a C file with the SQLite source code, and is meant to be compiled and then linked with `sqlite` to create a single binary that contains the benchmark driver and the application. To ensure that only the application is instrumented, preventing additional, unwanted overhead from the benchmark driver, we patched the driver's source code, adding a `nocf_check` attribute to each function. This attribute prevents the compiler from adding: (1) IBT or `FineIBT` instrumentation to function prologues, and (2) `FineIBT` instrumentation at indirect branch sites. Both IBT and `FineIBT` incur a negligible runtime slowdown (< 0.5%).

#### 6.1.3 Code-size Increase.
We measured the impact of `FineIBT`'s instrumentation on the code sections (i.e., sections marked with `SHF_ALLOC` and `SHF_EXECINSTR` flags) of the resulting ELF files, using the binaries of SPEC CPU2017 and those of the 4 real-world applications above, including all their DSO dependencies. Table 3 summarizes our findings. Unsurprisingly, IBT incurs a negligible code-size increase (< 1% on avg.), as it only requires instrumenting address-taken functions and PLT entries with `endbr` instructions (4 bytes). `FineIBT` results in a larger increase: in the case of the 4 real-world applications, the space overhead of `FineIBT` ranges between 7.71%–18.11%, whereas in SPEC CPU2017 it ranges between 2.27%–19.05%. On average, `FineIBT`-instrumented binaries were smaller than their Clang-CFI counterparts across the SPEC CPU2017 binaries, whose space overhead ranges between 2.13%–23.21%. Most large percentage increases in `FineIBT`- and Clang-CFI-instrumented binaries are mainly due to instrumentation getting added to smaller programs, where the size of the instrumentation relative to the size of the binary is considerable and thus more apparent. Notably, we also used the error-handling variant of `FineIBT` (§4.3.3), which represents the worst-case space overhead for `FineIBT` as the instrumentation includes additional instructions to setup and call an error-handling function.

**Table 3: Code-size increase for IBT, FineIBT, and Clang-CFI.**

| Application | Vanilla KB | IBT KB | %-Chg | FineIBT (§4.3) KB | %-Chg | Clang-CFI KB | %-Chg |
|---|---|---|---|---|---|---|---|
| **Redis (Total)** | **2787** | **2808** | **0.74%** | **3292** | **18.11%** | **–** | **–** |
| redis-server | 1890 | 1905 | 0.80% | 2248 | 18.90% | – | – |
| libc.so | 897 | 903 | 0.60% | 1045 | 16.44% | – | – |
| **SQLite (Total)** | **2237** | **2247** | **0.44%** | **2409** | **7.71%** | **–** | **–** |
| speedtest | 1340 | 1344 | 0.33% | 1364 | 1.86% | – | – |
| libc.so | 897 | 903 | 0.60% | 1045 | 16.44% | – | – |
| **Nginx (Total)** | **7005** | **7064** | **0.83%** | **7842** | **11.94%** | **–** | **–** |
| nginx | 1606 | 1619 | 0.81% | 1714 | 6.70% | – | – |
| libpcre.so | 576 | 577 | 0.11% | 582 | 0.97% | – | – |
| libssl.so | 649 | 660 | 1.67% | 726 | 11.83% | – | – |
| libcrypto.so | 2880 | 2905 | 0.86% | 3344 | 16.10% | – | – |
| libz.so | 130 | 132 | 0.95% | 144 | 10.04% | – | – |
| libGeoIP.so | 266 | 268 | 0.82% | 288 | 8.32% | – | – |
| libc.so | 897 | 903 | 0.60% | 1045 | 16.44% | – | – |
| **MariaDB (Total)** | **33422** | **33650** | **0.68%** | **37821** | **13.16%** | **–** | **–** |
| mariadbd | 26765 | 26924 | 0.60% | 30144 | 12.63% | – | – |
| libpcre2-8.so | 655 | 658 | 0.43% | 670 | 2.23% | – | – |
| libbz2.so | 76 | 77 | 1.09% | 81 | 6.62% | – | – |
| libaio.so | 4 | 4 | 4.94% | 6 | 61.03% | – | – |
| libz.so | 130 | 132 | 0.95% | 144 | 10.04% | – | – |
| libssl.so | 649 | 660 | 1.67% | 726 | 11.83% | – | – |
| libcrypto.so | 2880 | 2905 | 0.86% | 3344 | 16.10% | – | – |
| libc.so | 897 | 903 | 0.60% | 1045 | 16.44% | – | – |
| libc++.so | 987 | 1005 | 1.83% | 1248 | 26.52% | – | – |
| libc++abi.so | 327 | 330 | 0.84% | 354 | 8.31% | – | – |
| libunwind.so | 52 | 53 | 1.41% | 59 | 14.15% | – | – |
| **600.perlbench** | **3685** | **3696** | **0.30%** | **3865** | **4.89%** | **3928** | **6.59%** |
| **602.gcc** | **12082** | **12104** | **0.18%** | **12356** | **2.27%** | **12339** | **2.13%** |
| **605.mcf** | **923** | **929** | **0.63%** | **1072** | **16.15%** | **1131** | **22.59%** |
| **620.omnetpp** | **4244** | **4312** | **1.60%** | **4875** | **14.87%** | **5097** | **20.10%** |
| **623.xalancbmk** | **6805** | **6907** | **1.50%** | **7647** | **12.37%** | **8385** | **23.21%** |
| **625.x264** | **1588** | **1596** | **0.48%** | **1750** | **10.16%** | **1829** | **15.12%** |
| **631.deepsjeng** | **2390** | **2417** | **1.14%** | **2835** | **18.61%** | **2641** | **10.51%** |
| **641.leela** | **2366** | **2395** | **1.23%** | **2816** | **19.05%** | **2623** | **10.88%** |
| **657.xz** | **1059** | **1065** | **0.61%** | **1211** | **14.42%** | **1275** | **20.44%** |

**Table 4: ConFIRM [1] results.**

| Test | Result | Description |
|---|---|---|
| callback | ✓ | Callbacks support |
| code_coop | ✓ | COOP attack [151] resilience |
| convention | ✓ | Different x86 calling conv. support |
| cppeh | ✓ | C++ exception handling support |
| data_symbl | ✓ | Import/export data sym. handling |
| fptr | ✓ | Indirect function call support |
| jit | ✗ | Runtime-generated code support |
| load_time_dynlnk | ✓ | Load-time function resolution |
| mem | ✓ | Memory mgmt. API support |
| multithreading | ✓ | Concurrent thread exec. support |
| pic | ✓ | PIC/PIE support |
| ret | ✓ | Return-address validation |
| run_time_dynlnk | ✓ | Run-time function resolution |
| signal | ✓ | Signal handling support |
| switch | ✓ | switch-based CF support |
| tail_call | ✓ | Tail-call optimizations |
| unmatched_pair | ✓ | Unmatched call/ret pairs |
| vtbl_call | ✓ | Virtual function support |

**Table 5: IBT-instrumentation elision (NOPout) results.**

| Application | AT-elided | Pages | KB |
|---|---|---|---|
| redis-server | 858 (19.05%) | 206 | 844 |
| sqlite | 896 (34.07%) | 165 | 676 |
| nginx | 2873 (19.33%) | 606 | 2482 |
| mariadbd | 17291 (44.98%) | 1915 | 7844 |
| 600.perlbench | 942 (35.19%) | 246 | 1008 |
| 602.gcc | 3646 (67.71%) | 665 | 2724 |
| 605.mcf | 83 (4.57%) | 45 | 184 |
| 620.omnetpp | 5623 (71.03%) | 405 | 1659 |
| 623.xalancbmk | 8023 (77.78%) | 644 | 2638 |
| 625.x264 | 302 (14.82%) | 76 | 311 |
| 631.deepsjeng | 1690 (42.45%) | 189 | 774 |
| 641.leela | 1722 (42.91%) | 194 | 795 |
| 657.xz | 176 (9.17%) | 59 | 242 |

## 6.2 Security

*6.2.1 ConFIRM.* We evaluated the effectiveness (i.e., security) and compatibility of FineIBT using the Linux-related tests [1] from the ConFIRM CFI benchmarking suite [180]. Table 4 summarizes our findings. FineIBT successfully passed all (Linux) tests with the exception of jit, which involved an indirect branch to JIT-compiled code. FineIBT is compiler-based, and hence the respective code was not instrumented with FineIBT IRMs—a problem easily resolved by making JIT engines FineIBT-aware.

Moreover, while FineIBT passed the compatibility check for the multithreading test, running it without issues, FineIBT failed the security check. To test security, multithreading spawns attacker and victim threads and attempts to have the attacker overwrite a return address of the victim thread. Since FineIBT does not provide return address protection the attack succeeds unless FineIBT is paired with a backward-edge defense such as Intel CET's shadow stack. Lastly, recall that FineIBT does not propose nor rely on a specific CFI policy; FineIBT is an *enforcement mechanism* for a wide range of CFI policies, which allow for different security vs. compatibility trade-offs. Earlier works have extensively investigated the security of different CFI policies (coarse- vs. fine- vs. finer-grain) and hence we do not attempt to repeat them here (see §2.2).

*6.2.2 NOPout.* We applied NOPout on the four real-world applications (§6.1.2), as well as on the 9 C/C++ benchmarks of the SPEC-speed 2017 Integer suite (§6.1.1), including all their DSO dependencies, to assess the impact of endbr elision (§4.5.1). (NOPout removes endbr instructions, via code patching, resulting in an address space instantiation where only the minimum set of address-taken functions are protected for the application to function properly.)

Table 5 summarizes our findings. Col. 2 corresponds to the number of address-taken (AT) functions that are removed from the set(s) of allowed branch targets, post load-time. The percentages in parentheses indicate by how much the total set of endbr-protected functions was reduced, or, in other words, how much more stringent/secure the enforced CFI policy will be. Col. 3 shows the number of 4KB pages that the respective AT functions span, while col. 4 outlines the memory overhead that NOPout incurs due to code patching: i.e., when patching shared code pages, COW (copy-on-write) kicks in, effectively duplicating the target page(s). The additional memory required by NOPout ranges between 184KB–7844KB for reducing the valid targets by 83–17291 functions, respectively.

```
1   main:                          /* caller */
2     ...
3     movz  w9, #0x3a, lsl #16     /* SID = 0x3a0000 */
4     blr   x0                     /* x0 = &func */
5     ...
6     bl    func_entry
7     ...
8   .func_finebti_coldpath:
9     ...                          /* arg0, ..., argn */
10    bl    __finebti_chk_fail@PLT
11  func:                          /* callee */
12    bti   c
13    subs  w9, w9, #0x3a0, lsl #12 /* SID=0x3a0000 */
14    bne   .func_finebti_coldpath
15  func_entry:
16    ...
```

**Listing 4: FineIBT IRM code ported to ARM.**

Currently, NOPout patches every required code page *eagerly*, right after the loading process is completed (§4.5.1). In the future, we plan on performing the patching *lazily*, as code pages are brought into the page cache (via #PF exceptions), minimizing NOPout's memory cost(s).

## 7 DISCUSSION

### 7.1 FineIBT on ARM-based Platforms

While the design of FineIBT (see §4) largely focuses on x86-64, FineIBT is not limited to this architecture only. The concept(s) behind FineIBT can be applied to any hardware-assisted, coarse-grain, forward-edge CFI scheme that is similar to IBT.

ARM BTI (Branch Target Identification) is a CFI hardware feature of AArch64, introduced in ARMv8.5 [98], which is analogous to Intel IBT. It mandates that indirect branches must target a specific instruction, namely bti, else a fault occurs. BTI does differ slightly from IBT in that the bti instruction encodes an indirect branch *type* as an operand, allowing only certain indirect control-flows to target it; e.g., only indirect calls, but not indirect jumps, can target bti c instructions; and vice versa for bti j.

Since BTI provides similar (control-flow) guarantees to IBT, through the same mechanism (i.e., a "landing pad" instruction, bti), we can easily port the IRM logic of FineIBT to ARM, as shown in Listing 4: that is, by replacing each x86-64 instruction in the FineIBT IRM code (see Listing 2) with an equivalent A64 instruction. However, due to instructions being fixed-length (4 bytes each) in AArch64, the immediate-value operands needed for SID checking are limited. For example, the movz instruction (ln. 3) only allows 16-bit immediate values, shifted left by an optional 16-bit offset[1], to be loaded onto a register. Similarly, the subs instruction (ln. 13) used for SID comparison only allows subtraction with a 12-bit immediate value (optionally left-shifted by 12-bits). To get a quantity of unique SIDs equivalent to x86-64, additional instructions are required, which we leave for future work to determine if necessary.

### 7.2 Resilience against Spectre

Indirect branch prediction can result in instructions getting transiently executed at an incorrect (i.e., predicted) target; this can be used by attacks such as Spectre-v2 (branch target injection). Notably, IBT partially mitigates these attacks by preventing or limiting transient execution at indirect branch targets without an endbr (§2.3.3). FineIBT can provide additional protection against such attacks by limiting speculation at indirect targets for invalid control flows (i.e., those which fail the SID check) both in the architectural and speculative domains. However, the IRM code of FineIBT (§4.2, §4.3) contains a conditional branch (je in Listing 1, jne in Listing 2) that can be targeted by a Spectre-v1 attack [89], whereby the CPU speculates beyond the SID check, transiently executing subsequent instructions (i.e., function code under the *_entry label) when there is a CFI violation.

To investigate the feasibility of the above threat, we mounted a Spectre-v1 attack on FineIBT (on par with our threat model) to measure: (1) the *likelihood* of transiently executing instructions after the SID check; and (2) the *size* of the speculation window (i.e., the number of transiently executed instructions) following the SID check. Our attack is modeled after the following scenario. First, an adversary provides input to the victim program to repeatedly invoke a function indirectly, training the conditional branch predictor to anticipate that FineIBT's SID check will always pass. Second, the adversary overwrites a (different) function pointer with the address of the aforementioned (trained) function. Finally, the attacker causes the invocation of the overwritten function pointer, which will potentially result in the transient execution of the instructions following FineIBT's conditional branch, until the branch direction and target are resolved, ultimately crashing the program as the SID check fails architecturally. Depending on the transiently executed instructions, the attacker may be able to leak information from the process through micro-architectural side-channels [79].

A detailed description of our attack is available in Appendix A.4; Table 6 (Appendix A.4) summarizes our findings. The results of the experiment suggest an extremely low success rate, regardless of whether the conditional branch is a je (Listing 1) or jne (Listing 2). Without any nop instructions padding the Spectre gadget, we observed its transient execution 17 out of 10M attempts for both je- and jne-based IRMs. With ≈14 nops of padding, this number dropped to 0, and thus we estimate the speculative window size in this scenario to be ≈15 bytes (13b of nops plus 2b for the mov).

Although this speculative window could theoretically still be enough to leak information if combined with techniques such as SMT contention [117], our results indicate that FineIBT imposes meaningful restrictions on potentially-exploitable gadgets that occur after FineIBT's prologue instrumentation, even in the face of speculative CFI violations. Thus, we contend that FineIBT reduces the speculative attack surface of a program.

### 7.3 Adoption

The release of Linux kernel v6.2 added a version of FineIBT tailored for the kernel in addition to enabling kernel IBT by default on supported (i.e., x86) platforms [107]. While this bodes well for the adoption of the userland form of FineIBT presented in this paper, it does not directly speak to the most significant barrier to adoption: the ABI changes required to securely support cross-DSO function calls. We believe our proposed ABI changes will not impede FineIBT's adoption for three main reasons. First, our changes are minimal when considered in tandem with the existing ABI changes that were introduced to support IBT. With the IBT PLT as a starting point, we merely add SID-checking code and rearrange the existing instructions across three ELF sections—two sections were already used by IBT PLT; we just added one more. Second, ABI changes regarding the PLT are not uncommon: the ABI was recently amended to accommodate the aforementioned IBT PLT, as well as Intel's Memory Protection Extensions (MPX) hardware feature [21]. Third, the behavior of the FineIBT PLT is the same as the standard PLT, supporting both eager and lazy binding transparently (i.e., the target of the cross-DSO function call need not be instrumented with FineIBT).

---

[1]32- and 48-bit left-shift is also available, if the 64-bit instruction variant is used.

### 7.4 Limitations and Future Work

*7.4.1 Context Sensitivity.* The current design of the FineIBT IRM code only considers the origin of the most recent indirect branch when deciding whether control flow is legal—i.e., a decision is made based on a single SID loaded into a register by the caller. Considering a history of multiple indirect branches can further improve the effectiveness of the scheme for a given policy by providing more context for decision-making [166]. In order to record additional information, making FineIBT context-aware, the following changes can be made to our existing design: (1) SID values can be shortened to allow storage of multiple SIDs in a register—e.g., two 4-byte SIDs ↝ eight 1-byte SIDs—; (2) the register that holds SIDs can be preserved to prevent non-FineIBT functionality from trashing the saved context; (3) the caller-side FineIBT IRM can be modified to store multiple SIDs in a register, similar to our FineIBT PLT instrumentation that handles delayed binding (§4.4.1); and (4) the callee-side instrumentation can be modified to verify multiple SIDs. Note that for (4), the worst case would require unpacking the SID register and individually verifying each stored SID, however, in the best case, multiple SIDs can be validated at once. For example, if SIDs are 1-byte, and a given target can only be reached by a constant set of eight indirect branches, the target's instrumentation can perform a single, 64-bit comparison to validate the prior eight branches instead of eight individual comparisons, one for each SID.

*7.4.2 Adaptive CFI.* FineIBT is agnostic to the policy it enforces (§4.1.2), which means its effectiveness is only as good as the enforced policy. If FineIBT is made to enforce a coarse-grain policy, it may be susceptible to exploitation [37, 57, 73, 74], and similarly, an imprecise fine-grain policy may also fall victim to attacks [36, 66, 151]. To combat this, future work can build upon FineIBT to create adaptive CFI schemes whereby effectiveness is continually refined as a hardened program runs. IBT (or BTI) can provide coarse-grain protection, anchoring indirect control flows to targets in a single, large equivalence class, and the FineIBT IRM code can be used to separate-out multiple equivalence classes which are further pruned using information available at runtime. Ideally, such a scheme would allow a program to start enforcing any policy granularity and eventually converge on a policy that restricts every indirect control flow transfer to a single target (e.g., UCT [82]).

*7.4.3 Hardware Improvements.* As discussed in Section 7.1, ARM BTI encodes more information than Intel IBT. Specifically, the bti instruction differentiates between different types of indirect branches, offering a small effectiveness boost and improving the granularity of the base mechanism over IBT. Future work may offer similar capabilities to IBT, or, more significantly, implement the FineIBT IRM code entirely in hardware. This would reduce the already minimal overhead of FineIBT even further while maintaining strong effectiveness guarantees as demonstrated by prior work [42, 55, 96, 133, 158].

## 8 CONCLUSION

We studied the design, implementation, and evaluation of FineIBT: a CFI enforcement mechanism for improving the precision of hardware-based CFI solutions, like Intel IBT (and ARM BTI), by instrumenting program code with performant IRMs that reduce the valid/allowed targets of indirect forward-edge transfers. We studied the design of FineIBT on the x86-64 architecture, and implemented and evaluated it on Linux and the LLVM toolchain. We optimized FineIBT's IRM code to be compact and incur low runtime and memory overheads, but at the same time be generic, so as to support a range of different CFI policies. FineIBT incurs negligible runtime slowdowns (≈0%–1.94% in SPEC CPU2017 and ≈0%–1.92% in real-world applications) outperforming Clang-CFI. In addition, we investigated the effectiveness/security and compatibility of FineIBT using the ConFIRM CFI benchmarking suite, demonstrating that FineIBT's IRMs provide complete coverage in the presence of modern software features, while supporting a wide range of CFI policies (coarse- vs. fine- vs. finer-grain) with the same, predictable performance behavior.

### Availability

The prototype implementation of FineIBT is available at:
https://gitlab.com/brown-ssl/fineibt

### ACKNOWLEDGMENTS

We thank our anonymous shepherd and reviewers for their valuable feedback. This work was supported in part by the CIFellows 2020 program, through award CIF2020-BU-04, and the National Science Foundation (NSF), through award CNS-2238467. Any opinions, findings, and conclusions or recommendations expressed herein are those of the authors and do not necessarily reflect the views of the US government, NSF, CRA, or Intel.

# A APPENDIX

## A.1 Clang-CFI Instrumentation

Listing 5 illustrates how Clang-CFI confines forward-edge transfers (backward-edges are not covered) in x86-64. Functions `0x4011d0` and `0x4011f0` (originally `f0` and `f1`) are address-taken, while `func` (`0x401140`) contains an indirect control-flow transfer (ln. 9).

Clang-CFI first identifies all address-taken functions, and places them into equivalence classes (i.e., sets) based on their type signatures—i.e., two (address-taken) functions belong to the same set, *iff* their type signatures match exactly [43]. In the example above, `0x4011d0` and `0x4011f0` belong to the same set. Next, the original symbols of address-taken functions are mangled, by appending the `.cfi`

```
1  0x401140 <func>:
2    ...
3    40116a:  mov    $0x401230,%ecx
4    40116f:  mov    %rax,%rdx
5    401172:  sub    %rcx,%rdx
6    401175:  rol    $0x3d,%rdx
7    401179:  cmp    $0x2,%rdx
8    40117d:  jae    4011c5
9    40117f:  callq  *%rax
10   ...
11   4011a3:  callq  4011d0 <f0.cfi>
12   ...
13   4011c5:  ud2
14   ...
15
16  0x4011d0 <f0.cfi>:
17   ...
18
19  0x4011f0 <f1.cfi>:
20   ...
21
22  0x401250 <f0>:
23   401250:  jmpq   4011d0 <f0.cfi>
24   401255:  int3
25   401256:  int3
26   401257:  int3
27  0x401258 <f1>:
28   401258:  jmpq   4011f0 <f1.cfi>
29   40125d:  int3
30   40125e:  int3
31   40125f:  int3
32   ...
```

**Listing 5: Clang-CFI IRM example (x86-64).**

suffix to them (e.g., `f0 ↝ f0.cfi`). In addition, each set is accompanied by a trampoline-like construct, which contains an entry for every function in the set. In above example, the trampoline that covers `0x4011d0` and `0x4011f0` spans ln. 23–ln. 31, and contains two entries. Each such entry (ln. 23–ln. 26 and ln. 28–ln. 31) is 8-byte aligned, and consists of a direct branch to its corresponding address-taken function (ln. 16, ln. 19). In x86, this branch is implemented via a direct `jmp` instruction, which may span up to 5 bytes (including the operand/branch target); the remaining bytes of each "slot" are filled with (single-byte) `int3` instructions to trap accesses to them (ln. 24–ln. 26, ln. 29–ln. 31). Lastly, Clang-CFI associates each trampoline slot with the symbol name of the original, address-taken function that corresponds to it. In other words, `0x4011d0` (originally function `f0`) is renamed to `f0.cfi`, and an entry for it is created at `0x401250` (trampoline), which is now associated with symbol `f0`. (The process is identical for function `f1`.) The net effect of the above is that whenever the address of `f0` or `f1` is taken, `0x401250` or `0x401258` will now be returned, instead of `0x4011d0` and `0x4011f0`, effectively "forcing" control paths that indirectly reach `f0`/`f1` to do so via the trampoline code.

```
1   PLT0:  pushq    GOT+8(%rip)      /*  GOT[1]  */
2          jmp      *GOT+16(%rip)    /*  GOT[2]  */
3          nopl     0x0(%rax)        /*   PAD    */
4   ...
5   PLT3:  jmp      *fsym3@GOT(%rip) /*  GOT[5]  */
6          pushq    $0x2
7          jmp      PLT0
8   PLT4:  jmp      *fsym4@GOT(%rip) /*  GOT[6]  */
9          pushq    $0x3
10         jmp      PLT0
11  ...
12  PLTn:  jmp      *fsymn@GOT(%rip) /* GOT[n+2] */
13         pushq    $0xn-1
14         jmp      PLT0
```

**Listing 6: Example of a non-IBT PLT (x86-64 ABI).**

Direct invocations of f0/f1 use the `.cfi` symbols instead (ln. 11). Call sites that correspond to indirect control-flow transfers are further instrumented with an IRM that performs confinement as follows. First, the type of the involved pointer that is dereferenced (e.g., function pointer, vtable entry) is analyzed, and matched with one of the sets that correspond to equivalence classes. In the example above, a function-pointer dereference occurs in ln. 9, and the target of the respective indirect branch in loaded in the general-purpose register %rax. Moreover, the type of that function pointer matches the type of functions `0x4011d0` and `0x4011f0` (originally f0 and f1), and so it is further associated with the set (equivalence class) whose trampoline starts at address `0x401250`. The IRM that confines the indirect branch instruction in ln. 9 spans ln. 3–ln. 8 (plus ln. 13), and effectively asserts if the value in %rax points to an address that corresponds to its associated trampoline. That is, if the value of %rax indeed points within `0x401250`–`0x40125f`, then the computed control-flow transfer (ln. 9) will be allowed, ultimately reaching `0x4011d0` (f0.cfi) or `0x4011f0` (f1.cfi) via the surrogate symbols at `0x401250` (f0) and `0x401258` (f1), respectively.

### A.2 Non-IBT PLT

In Linux, the x86-64 platform leverages the ELF (Executable and Linkable Format) binary/file format for storing executables, object files, dynamic shared objects (DSOs), and even core dumps [99]. Cross-DSO function calls correspond to control-flow transfers from one DSO (e.g., shared library; `.so` ELF file of type ET_DYN), or the main application executable (ELF file of type ET_EXEC), to functions (i.e., symbols of type FUNC) that are located in different DSOs. Given that the final placement of the corresponding DSOs, in the virtual address space, will be performed at load time, by the dynamic linker/loader (ld.so)—and is affected by features like ASLR—, special care is required for supporting cross-DSO function calls.

The x86-64 ABI [21] specifies how this process is performed and what ELF features it entails. Cross-DSO function calls leverage the PLT/GOT (Procedure Linkage Table/Global Offset Table) mechanism [131]. Every DSO that invokes functions that are "external" to it (i.e., they belong to a different DSO) contains a special (ELF) code section dubbed `.plt`.

This section is further split into "slots" (i.e., entries), of 16 bytes each, as shown in Listing 6. The first entry (PLT0) is a special one, while each of the rest (PLT1–PLTn) correspond to external function calls. Whenever an ELF-compatible toolchain (e.g., GCC, LLVM) needs to emit code for an external function call, it will do so by generating a (direct) `call` instruction that transfers control to the respective PLT slot. For instance, in the example above, if the DSO whose PLT is shown in Listing 6, needs to invoke function fsym3, which is located in a different DSO, then the toolchain will generate call sites that transfer control (directly) to PLT3. (The process is similar for fsym4–fsymn.) Every PLT slot is associated with an entry in GOT—which, in turn, is stored into a special (ELF) data section dubbed `.got`. GOT is effectively an "array" of memory addresses, each for every external symbol (of type FUNC, OBJECT, *etc.*) used by the respective DSO. In other words, only a subset of the GOT entries are related to PLT slots; the rest correspond to (cross-DSO) data-object references, and are initialized eagerly.

If, say, the DSO whose PLT is shown in Listing 6 wants to invoke fsym3, which is located in a different DSO, the process is as follows: first, the control will be transferred to PLT3 (ln. 5), by a direct `call` instruction (located in one of the executable sections, e.g., `.text`) of the current DSO; assuming that "PLT[3]" is associated with "GOT[5]", the memory-indirect `jmp` instruction in ln. 5 will transfer control to the code location stored at GOT[5] (fsym3@GOT); initially, at load time, ld.so makes GOT[5] point at the push instruction of PLT3 (ln. 6)—and GOT[6] at the push instruction of PLT4 (ln. 9), *etc.*—effectively causing the indirect `jmp` instructions (ln. 5, ln. 8, ln. 12, . . .) to "fall-through;" next, an index (non-negative) value is pushed onto the stack (ln. 6), and the control is (directly) transferred to PLT0 (ln. 7); PLT[0] is a special entry, as it pushes onto the stack a reference to the current DSO (stored at GOT[1], ln. 1) and then transfers control to the address stored at GOT[2], via the memory-indirect `jmp` instruction in ln. 2; GOT[2] is also initialized during load time (by ld.so), and contains the address of the (part of the) dynamic linker/loader for performing symbol resolution; at this point, ld.so will unwind the stack, identify that the binding request came from the current DSO, via the value pushed onto the stack in ln. 1, as well as the symbol in question (i.e., fsym3), via the index pushed onto the stack in ln. 6, and look for an already-loaded (and relocated) DSO (i.e., the foreign DSO) that provides (or "exports") fsym3 in its `.dynsym` ELF section; once the address of the symbol, in the foreign DSO, is identified, ld.so will finally store it at GOT[5], and control will be transferred to fsym3. Next time a direct `call` instruction transfers control to PLT3, the memory-indirect `jmp` instruction in ln. 5 will branch straight to fsym3 (via GOT[5]), without triggering the symbol resolution process again.

This intertwined mechanism is called lazy binding as it results in resolving the symbol in question (e.g., function fsym3), dynamically, at runtime, upon the first use. However, note that only code symbols (of type FUNC) can be resolved lazily; data symbols (e.g., of type OBJECT) are always resolved eagerly, at load time (by ld.so). RELRO (relocation read-only) [147] builds upon this observation and further separates the PLT-related part(s) of GOT, placing the respective entries under a different ELF section dubbed `.got.plt`; the rest of the GOT entries (which are resolved eagerly) are left under `.got`, which is write-protected by the dynamic linker/loader once load-time (data) symbol resolution is completed. RELRO (`-z relro`

```
1   PLT0:  pushq    GOT+8(%rip)      /*  GOT[1]  */
2          jmp      *GOT+16(%rip)    /*  GOT[2]  */
3          nopl     0x0(%rax)        /*  PAD     */
4   ...
5   PLT3:  endbr64
6          pushq    $0x2
7          jmp      PLT0
8          xchg     %ax,%ax          /*  PAD     */
9   PLT4:  endbr64
10         pushq    $0x3
11         jmp      PLT0
12         xchg     %ax,%ax          /*  PAD     */
13  ...
14  PLTn:  endbr64
15         pushq    $0xn-1
16         jmp      PLT0
17         xchg     %ax,%ax          /*  PAD     */
18
19  ...
20  SPLT3: endbr64
21         jmp      *fsym3@GOT(%rip) /*  GOT[5]  */
22         nopw     0x0(%rax,%rax,1) /*  PAD     */
23  SPLT4: endbr64
24         jmp      *fsym4@GOT(%rip) /*  GOT[6]  */
25         nopw     0x0(%rax,%rax,1) /*  PAD     */
26  ...
27  SPLTn: endbr64
28         jmp      *fsymn@GOT(%rip) /* GOT[n+2] */
29         nopw     0x0(%rax,%rax,1) /*  PAD     */
```

**Listing 7: IBT PLT (x86-64 ABI).**

or `-Wl,-z,relro`; denoted with the `PT_GNU_RELRO` ELF header) aims at reducing the attack surface of GOT, effectively making the entries immutable that should remain constant post load-time. Lastly, in addition to the above, RELRO can be configured to operate in "full" mode (i.e., by enabling `BIND_NOW`; `-z now` or `-Wl,-z,now`), which results in turning on eager binding. Under full RELRO, the dynamic linker/loader resolves *all* GOT entries at load time and write-protects `.got` (under full RELRO, `.got.plt` is again merged with `.got`, as there is no reason to keep them separate), thereby preventing an attacker from tampering with the GOT completely (and the binding process, in general). Note that under full RELRO, PLT entries branch straight to symbols in foreign DSOs, skipping the on-demand resolution process via PLT[0].

### A.3  IBT PLT

The main differences with the original, plain x86-64 PLT are as follows. First, the PLT of each IBT-hardened DSO consists of *two* "tables," mapped at the ELF sections `.plt` and `.plt.sec`, respectively. Whenever an IBT-compatible toolchain (e.g., GCC, LLVM) needs to emit code for an external function call (i.e., a control-flow transfer to a symbol of type FUNC that is located in a foreign DSO), it does so by emitting a (direct) `call` instruction that links-with an entry in `.plt.sec` (SPLT1–SPLTn in Listing 7).

For example, if the current DSO needs to, say, invoke the (code) symbol fsym3, which is located in a different `.so` ELF file, then the respective (direct) `call` will be in the following form: `call fsym3@SPLT`—i.e., a direct branch to symbol SPLT3 (ln.20). (`.plt.sec` includes an entry for each external symbol of type FUNC.) Next, at runtime, whenever control reaches SPLT3, the memory-indirect `jmp` instruction in ln.21 will transfer control either to (a) fsym3 at a foreign DSO (*iff* symbol resolution has already been performed by the dynamic linker/loader, `ld.so`), or (b) PLT3 (ln.5) at the `.plt` section of the current DSO, via the corresponding GOT entry/-slot (i.e., fsym3@GOT, "GOT[5]"). (The process is similar for fsym4–fsymn.) Lastly, every entry in `.plt.sec` has an associated entry in `.plt` (SPLTn ⤳ PLTn; ln.5, ln.9, …, ln.14) for handling lazy (or delayed) binding as usual.

Note that *every* entry in `.plt.sec` begins with an endbr instruction (SPLT3; ln.20, SPLT4; ln.23, …, SPLTn; ln.27), in order to support address-taken PLT slots—i.e., allow PLT entries to be targeted by indirect `call` instructions. (This typically occurs when a symbol with external linkage is address-taken in the current DSO.) In addition, every entry in `.plt` needs also to begin with an endbr instruction, as they are targeted by the corresponding memory-indirect `jmp` instructions in `.plt.sec`, which, in turn, leverage the untrusted GOT—recall that all PLT-related GOT slots are necessarily writeable (as they need to be updated, lazily, by `ld.so`) and hence an attacker can tamper with them.

This approach increases the set of allowed branch targets by a great margin: (1) every address-taken function is naturally an allowed target (i.e., endbr-instrumented), followed by (2) every symbol with non-local linkage, in every DSO (as they can be targets of the memory-indirect `jmp` instructions in `.plt.sec` sections), followed by (3) every `.plt.sec` slot (they can be address-taken), followed by (4) every `.plt` slot (targets of `.plt.sec` entries).

### A.4  Resilience against Spectre-v1

We implement the attack described in Section 7.2 with a tight loop that is run 10 million times. In each iteration, we indirectly branch to a FineIBT instrumented prologue 10 times with a valid SID, to train the conditional branch predictor, before indirectly branching to the prologue with an invalid SID. We detect misspeculation during the invalid SID check with a memory-read Spectre gadget (`movl (%rbx), %eax`) that proceeds the check whose transient execution we observe with a Flush+Reload [181] cache-based side-channel. This allows us to measure the likelihood of transiently executing instructions after the SID check. To determine the size of the speculative window, we pad the memory read with (single-byte) nop instructions, until we no longer observe the gadget executed (successfully) in a transient fashion. The assembly code of the aforementioned experiment is shown in Listing 8.

**Table 6: Speculative window size of the conditional branch in `FineIBT`'s IRM code.**

| Window Size (Bytes) | 2 | 3 | 4 | 5 | 6 | 7 | 8 | 9 | 10 | 11 | 12 | 13 | 14 | 15 | 16 | 17 |
|---|---|---|---|---|---|---|---|---|---|---|---|---|---|---|---|---|
| Hits/10M (jne) | 17 | 12 | 11 | 6 | 7 | 18 | 17 | 18 | 13 | 1 | 0 | 1 | 2 | 1 | 0 | 0 |
| Hits/10M (je) | 17 | 13 | 10 | 16 | 10 | 16 | 14 | 11 | 3 | 15 | 3 | 2 | 15 | 0 | 0 | 0 |

```
1    lea .fineibt_prologue(%rip), %r10
2    lea .do_test(%rip), %r11
3    lea .do_train(%rip), %r12
4    mov $0xa, %ecx /* init loop counter */
5  .loop_start:
6    clflush 0(%rbx) /* flush target cache line */
7    mfence
8    lfence
9    test %ecx, %ecx
10   cmove %r11, %r12 /* if ZF==1 do test run */
11   notrack jmp *%r12
12   nop
13 .do_train:
14   mov $0xdeadbeef, %eax /* valid SID */
15   jmp *%r10
16 .do_test:
17   mov $0xdeadbe42, %eax /* invalid SID */
18   jmp *%r10
19   nop
20 .fineibt_coldpath:
21   jmp .reload
22 .fineibt_prologue:
23   endbr64
24   sub $0xdeadbeef, %eax
25   jne .fineibt_coldpath
26   /* {0, ..., N} nops to measure window size */
27   movl (%rbx), %eax /* access memory */
28   dec %ecx
29   jmp .loop_start
30 .reload:
31   mfence
32   lfence
33   rdtsc /* start memory access timer */
34   lfence
35   movl %eax, %esi
36   movl (%rbx), %eax /* access memory */
37   lfence
38   rdtsc /* stop memory access timer */
39   subl %esi, %eax
```

**Listing 8: Assembly code that is used to measure the speculative behavior (i.e., resilience against Spectre-v1) of the conditional branch (`je`/`jne`) in `FineIBT`'s IRM code.**